\RequirePackage{lineno}
\documentclass[10pt,aps,prd,twocolumn,linenumbers,groupedaddress,floatfix,a4paper]{revtex4}
\usepackage{graphicx,amsmath,amssymb}
\usepackage{dcolumn}% Align table columns on decimal point
\usepackage{bm}% bold math
\usepackage{longtable}
\usepackage{color}
\usepackage{float}
\usepackage[english]{babel}
\begin{document}
%\linenumbers

%\title{Electrical potential and power of thundercloud measured by the GRAPES-3 experiment.}
%\title{Electrical properties of a thundercloud measured by muon tomography by the GRAPES-3 experiment.}
\title{Measurement of the electrical properties of a thundercloud through muon imaging by the GRAPES-3 experiment.}

% author's list
\author{B.~Hariharan}
\affiliation{Tata Institute of Fundamental Research, Homi Bhabha Road,
Mumbai 400005, India}
\altaffiliation{The GRAPES-3 Experiment, Cosmic Ray Laboratory, \\ Raj Bhavan, Ooty 643001, India}

\author{A.~Chandra}
\affiliation{Tata Institute of Fundamental Research, Homi Bhabha Road,
Mumbai 400005, India}
\altaffiliation{The GRAPES-3 Experiment, Cosmic Ray Laboratory, \\ Raj Bhavan, Ooty 643001, India}

\author{S.R.~Dugad}
\affiliation{Tata Institute of Fundamental Research, Homi Bhabha Road,
Mumbai 400005, India}
\altaffiliation{The GRAPES-3 Experiment, Cosmic Ray Laboratory, \\ Raj Bhavan, Ooty 643001, India}

\author{S.K.~Gupta}
%\email[]{gupta@grapes.tifr.res.in}
%\vskip -0.1in
\email[]{gupta.crl@gmail.com}
\affiliation{Tata Institute of Fundamental Research, Homi Bhabha Road,
Mumbai 400005, India}
%\vskip -0.1in
\altaffiliation{The GRAPES-3 Experiment, Cosmic Ray Laboratory, \\ Raj Bhavan, Ooty 643001, India}

\author{P.~Jagadeesan}
\affiliation{Tata Institute of Fundamental Research, Homi Bhabha Road,
Mumbai 400005, India}
\altaffiliation{The GRAPES-3 Experiment, Cosmic Ray Laboratory, \\ Raj Bhavan, Ooty 643001, India}

\author{A.~Jain}
\affiliation{Tata Institute of Fundamental Research, Homi Bhabha Road,
Mumbai 400005, India}
\altaffiliation{The GRAPES-3 Experiment, Cosmic Ray Laboratory, \\ Raj Bhavan, Ooty 643001, India}

\author{P.K.~Mohanty}
\affiliation{Tata Institute of Fundamental Research, Homi Bhabha Road, 
Mumbai 400005, India}
\altaffiliation{The GRAPES-3 Experiment, Cosmic Ray Laboratory, \\ Raj Bhavan, Ooty 643001, India}

\author{S.D.~Morris}
\affiliation{Tata Institute of Fundamental Research, Homi Bhabha Road, 
Mumbai 400005, India}
\altaffiliation{The GRAPES-3 Experiment, Cosmic Ray Laboratory, \\ Raj Bhavan, Ooty 643001, India}

\author{P.K.~Nayak}
\affiliation{Tata Institute of Fundamental Research, Homi Bhabha Road, 
Mumbai 400005, India}
\altaffiliation{The GRAPES-3 Experiment, Cosmic Ray Laboratory, \\ Raj Bhavan, Ooty 643001, India}

\author{P.S.~Rakshe}
\affiliation{Tata Institute of Fundamental Research, Homi Bhabha Road, 
Mumbai 400005, India}
\altaffiliation{The GRAPES-3 Experiment, Cosmic Ray Laboratory, \\ Raj Bhavan, Ooty 643001, India}

\author{K.~Ramesh}
\affiliation{Tata Institute of Fundamental Research, Homi Bhabha Road, 
Mumbai 400005, India}
\altaffiliation{The GRAPES-3 Experiment, Cosmic Ray Laboratory, \\ Raj Bhavan, Ooty 643001, India}

\author{B.S.~Rao}
\affiliation{Tata Institute of Fundamental Research, Homi Bhabha Road,
Mumbai 400005, India}
\altaffiliation{The GRAPES-3 Experiment, Cosmic Ray Laboratory, \\ Raj Bhavan, Ooty 643001, India}

\author{L.V.~Reddy}
\affiliation{Tata Institute of Fundamental Research, Homi Bhabha Road,
Mumbai 400005, India}
\altaffiliation{The GRAPES-3 Experiment, Cosmic Ray Laboratory, \\ Raj Bhavan, Ooty 643001, India}

\author{M.~Zuberi}
\affiliation{Tata Institute of Fundamental Research, Homi Bhabha Road,
Mumbai 400005, India}
\altaffiliation{The GRAPES-3 Experiment, Cosmic Ray Laboratory, \\ Raj Bhavan, Ooty 643001, India}

\author{Y.~Hayashi}
%\affiliation{Graduate School of Science, Osaka City University, 558-8585 Osaka, Japan}
\affiliation{Graduate School of Science, Osaka City University, Osaka, Japan}
\altaffiliation{The GRAPES-3 Experiment, Cosmic Ray Laboratory, \\ Raj Bhavan, Ooty 643001, India}

\author{S.~Kawakami}
%\affiliation{Graduate School of Science, Osaka City University, 558-8585 Osaka, Japan}
\affiliation{Graduate School of Science, Osaka City University, Osaka, Japan}
\altaffiliation{The GRAPES-3 Experiment, Cosmic Ray Laboratory, \\ Raj Bhavan, Ooty 643001, India}

\author{S.~Ahmad}
\affiliation{Aligarh Muslim University, Aligarh 202002, India}
\altaffiliation{The GRAPES-3 Experiment, Cosmic Ray Laboratory, \\ Raj Bhavan, Ooty 643001, India}

\author{H.~Kojima}
\affiliation{College of Engineering, Chubu University, Kasugai, Aichi, Japan}
%\affiliation{Faculty of Engineering, Aichi Institute of Technology, Toyota City, Japan}
\altaffiliation{The GRAPES-3 Experiment, Cosmic Ray Laboratory, \\ Raj Bhavan, Ooty 643001, India}

\author{A.~Oshima}
%\affiliation{College of Engineering, Chubu University, Kasugai, Aichi 487-8501, Japan}
\affiliation{College of Engineering, Chubu University, Kasugai, Aichi, Japan}
\altaffiliation{The GRAPES-3 Experiment, Cosmic Ray Laboratory, \\ Raj Bhavan, Ooty 643001, India}

\author{S.~Shibata}
%\affiliation{College of Engineering, Chubu University, Kasugai, Aichi 487-8501, Japan}
\affiliation{College of Engineering, Chubu University, Kasugai, Aichi, Japan}
\altaffiliation{The GRAPES-3 Experiment, Cosmic Ray Laboratory, \\ Raj Bhavan, Ooty 643001, India}

\author{Y.~Muraki}
\affiliation{Institute for Space-Earth Environmental Research, Nagoya University, Nagoya, Aichi Japan}
\altaffiliation{The GRAPES-3 Experiment, Cosmic Ray Laboratory, \\ Raj Bhavan, Ooty 643001, India}

\author{K.~Tanaka}
%\affiliation{Graduate School of Information Sciences, Hiroshima City University, Hiroshima 731-3194, Japan}
\affiliation{Graduate School of Information Sciences, Hiroshima City University, Hiroshima, Japan}
\altaffiliation{The GRAPES-3 Experiment, Cosmic Ray Laboratory, \\ Raj Bhavan, Ooty 643001, India}

\begin{abstract}
The GRAPES-3 muon telescope located in Ooty,
India records rapid ($\sim$10\,min) variations in the muon
intensity during major thunderstorms. Out of a total of 184
thunderstorms recorded during the interval April
2011--December 2014, the one on 1 December 2014 produced a
massive potential of 1.3\,GV. The electric field measured by
four well-separated (up to 6\,km) monitors on the ground was
used to help estimate some of the properties of this
thundercloud including its altitude and area that were found
to be 11.4\,km above mean sea level (amsl) and
$\geq$380\,km$^2$, respectively. A charging time of 6\,min
to reach 1.3\,GV implied the delivery of a power of
$\geq$2\,GW by this thundercloud that was moving at a
speed of $\sim$60\,km\,h$^{-1}$. This work possibly provides
the first direct evidence for the generation of GV potentials
in thunderclouds that could also possibly explain the
production of highest energy (100\,MeV) $\gamma$-rays in the
terrestrial $\gamma$-ray flashes.
\end{abstract}

\maketitle

\vskip -0.25in
Thunderstorms are a spectacular manifestation of the discharge
of massive electric potentials that develop in thunderclouds
during severe weather conditions. 
The first authoritative study of thunderstorms by Franklin
dates back to 1750s \cite{Franklin50}. A major advance in
their understanding occurred in 1920s when their dipole
structure was identified \cite{Wilson29}. However, actual
structure is more complex. The separation of electric
charges in thunderclouds occurs when supercooled
water-droplets make grazing contact with hail-pellets
(graupel) polarized by the fine-weather electric field
(120\,V\,m$^{-1}$) on Earth's surface. The rebounding
droplets acquire positive charge and are carried by
convective updraft toward the cloud-top while negatively
charged graupel fall toward cloud-base due to gravity.
This creates a vertical field that increases polarizing
charge on graupel thus, accelerating this process and
reinforcing vertical field, that grows exponentially
until air insulation breaks down and triggers a lightning
discharge. 
%Graupel falling at a speed of 30\,mm\,h$^{-1}$
%can produce vertical electric fields of 5000\,V\,cm$^{-1}$
%in about 10\,minutes
%involving the separation of $\sim$50\,C of charge, enough to
%initiate another lightning flash. The graupel movement sustains
%the process of producing a succession of lightning flashes at
%$\sim$30\,s intervals
\cite{Mason72}. Since the thickness of thunderclouds
extends to several kilometers, potentials of $\geq$1\,GV
could be generated \cite{Wilson29}.

A unique signature of massive electric potentials generated
in thunderclouds was the discovery of terrestrial
$\gamma$-ray flashes (TGFs) containing MeV photons by the
BATSE instrument aboard Compton $\gamma$-ray observatory.
The source of TGFs was identified to be thunderstorms in
the lower tropical atmosphere \cite{Fishman94}.
%Subsequently, the RHESSI data showed that $\gamma$-ray
%energies extended to 20\,MeV, produced through bremsstrahlung
%of electrons of yet higher energies \cite{Smith05}.
%The results from the $\gamma$-ray burst monitor aboard
%Fermi satellite extended the maximum $\gamma$-ray energy
%to 40\,MeV, implying the presence of $>$100\,MeV electrons
%\cite{Briggs10}. 
The detection of highest $\gamma$-ray
energy of 100\,MeV by the AGILE satellite would however,
require bremsstrahlung of very high-energy electrons and
presence of potentials of hundreds of MV \cite{Tavani11}.
The maximum thunderstorm potential measured in balloon
soundings is only 0.13\,GV \cite{Marshall01}, well short
of the magnitude needed to produce 100\,MeV
$\gamma$-rays \cite{Tavani11} and of 1\,GV predicted by
Wilson \cite{Wilson29}. MeV $\gamma$-rays produced in
thunderstorms have been detected on the ground, both
through triggered and natural lightening discharges,
showing a close connection of the TGFs detected from
space and from ground \cite{Dwyer03,Ringuette13}. Early
studies of the changes in muon intensity (I$_{\mu}$) at
low-energies (90\,MeV) were shown to be correlated
with the electric field of thunderstorms
\cite{Alexeenko87,Dorman03} and confirmed by the
results from Mt. Norikura \cite{Muraki04} and
elsewhere \cite{Chilingarian17}.

\begin{figure}[t]
\begin{center}
\includegraphics*[width=0.35\textwidth,angle=0,clip]{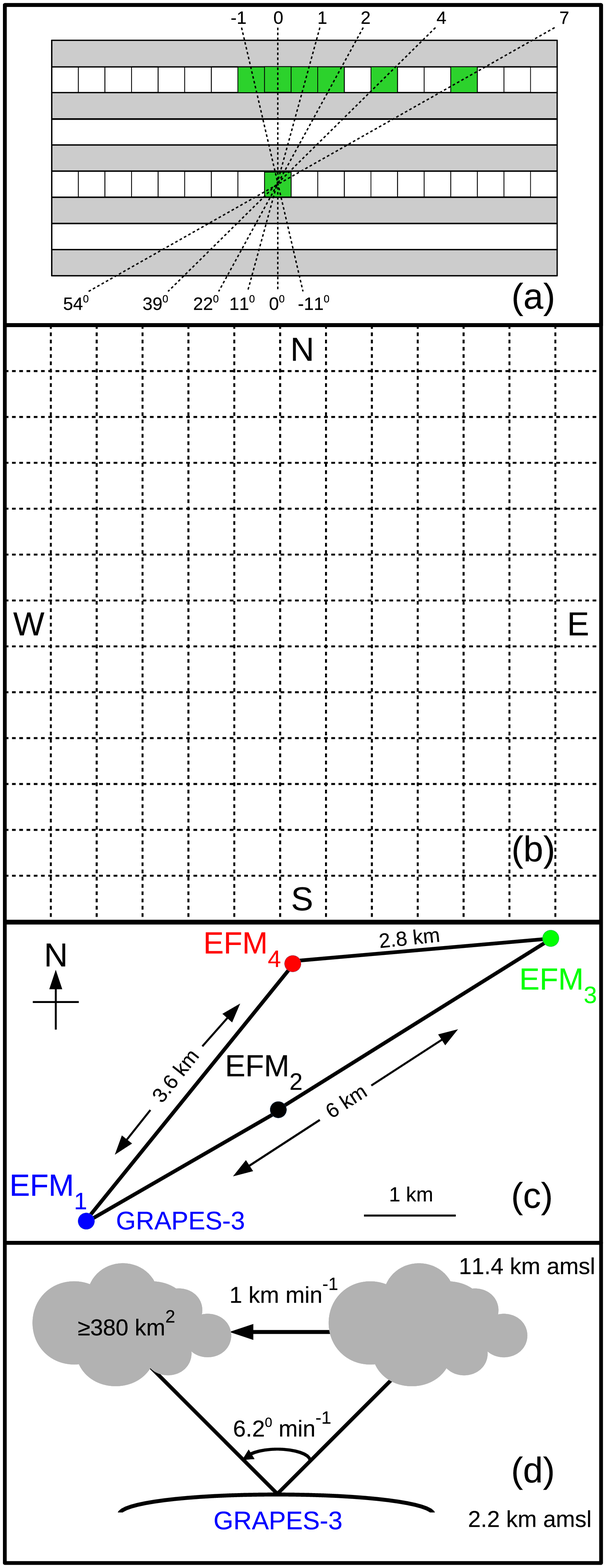}
\vskip -0.05in
\caption{\label{fig01} (a) Reconstruction of muon directions in
                       a single projection plane from PRC geometry,
                       (b) telescope field of view (FOV) of 2.3\,sr
                       segmented into 13$\times$13=169 directions,
                       (c) locations of EFMs labeled 1 to 4. 
                       Maximum distance of EFM$_1$ and
                       EFM$_3$\,=\,6\,km, (d) Schematic of 
                       thundercloud movement (linear and angular
                       velocities), altitude and area.}
\end{center}
\vskip -0.35in
\end{figure}
The GRAPES-3 muon telescope (G3MT) in Ooty (11.4$^\circ$N,
2200\,m\,amsl) studies astrophysics of cosmic rays (CRs)
through the measurement of I$_{\mu}$ produced by CRs. Its
detection element is a proportional counter (PRC) made from
steel pipes (6m$\times$0.1m$\times$0.1m). The G3MT consists
of 4 PRC layers under a 2\,m thick concrete-roof, resulting
in a threshold of E$_{\mu}$\,=\,1\,sec($\theta$)\,GeV, for
muons of zenith angle\,=\,$\theta$. This 4-layer
configuration enables muon reconstruction in two mutually
perpendicular planes and the two PRC layers in same
projection plane separated by $\sim$50\,cm permit muon
direction to be measured with $\sim$4$^{\circ}$ accuracy as
shown in Fig.\,\ref{fig01}a. Thus, the G3MT measures
I$_{\mu}$ in 169 directions over a field of view, hereafter
FOV\,=\,2.3\,sr as shown in Fig.\,\ref{fig01}b. Although,
the solid angle of 169 directions differ significantly,
but the area of thundercloud covered varies by only 19\%
\cite{Hari17}. Since $\sim$2.5$\times$10$^{6}$ muons are
recorded every minute, I$_{\mu}$ gets measured to 0.1\%
precision \cite{Gupta05,Mohanty16}. 

During thunderstorms, G3MT detects rapid changes
($\sim$10\,min) in I$_{\mu}$. Since the muon energies
exceed 1\,GeV, the presence of large electric potentials
is implied. To probe this phenomenon, electric field
monitors, hereafter ``EFM'' (Boltek model EFM-100
\cite{Boltek}) were installed in April 2011 at four
locations, at GRAPES-3, and three others a few km away
as shown in Fig.\,\ref{fig01}c. The data collected during
April 2011--December 2014 showed that 184 thunderstorms
were detected both by G3MT and EFMs. The seven largest
events with muon intensity variation
$\Delta$I$_{\mu} \geq$0.4\% were shortlisted. However,
except for the event on 1 December 2014 discussed here,
the EFM profiles of remaining six events were extremely
complex, that made association of $\Delta$I$_{\mu}$ and
electric field of a specific thundercloud difficult.

\begin{figure}[h]
\vskip -0.05in
\begin{center}
\includegraphics*[width=0.35\textwidth,angle=0,clip]{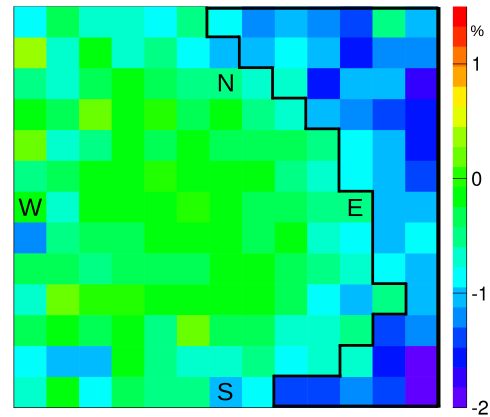}
\vskip -0.1in
\caption{\label{fig02} Muon intensity variation during 18\,min
                       thunderstorm. 45 out of 169, thunderstorm
                       affected contiguous directions are
                       enclosed by dark boundary. Color-coded
                       \% variation shown by a bar on right.
                       Thundercloud angular size in
                       N-S\,=\,74.6$^{\circ}$.}
\end{center}
\vskip -0.2in
\end{figure}
\begin{figure}[t]
\begin{center}
\includegraphics*[width=0.4\textwidth,angle=0,clip]{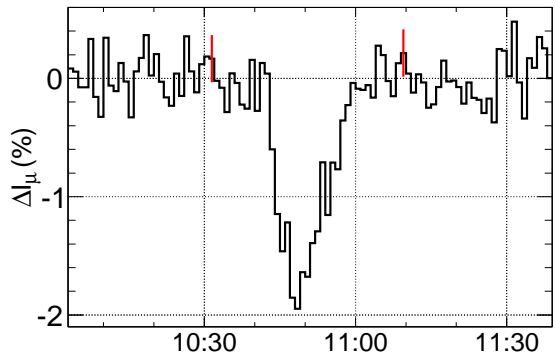}
\vskip -0.1in
\caption{\label{fig03} Maximum muon intensity variation
                       $\Delta$I$_{\mu}$\,=\,--2\%, starting
                       10:42\,UT, lasting 18\,min seen during
                       thunderstorm of 1 December 2014.
                       Vertical bars represent $\pm$\,1$\sigma$
                       errors.}
\end{center}
\vskip -0.15in
\end{figure}
\begin{figure}[h]
\vskip -0.05in
\begin{center}
\includegraphics*[width=0.4\textwidth,angle=0,clip]{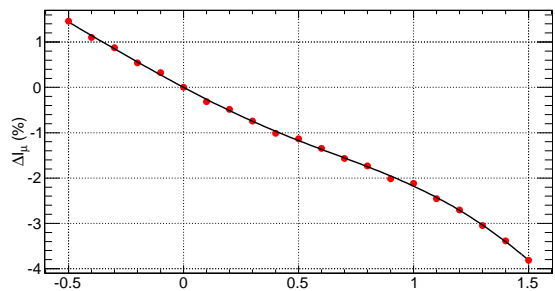}
\vskip -0.1in
\caption{\label{fig04} Dependence of $\Delta$I$_{\mu}$ on electric
                       potential (GV) across atmospheric layer
                       8--10\,km amsl, based on simulations for
                       45 directions shown in Fig.\,\ref{fig02}.} 
\end{center}
\vskip -0.2in
\end{figure}
\begin{figure}[h]
\begin{center}
\includegraphics*[width=0.4\textwidth,angle=0,clip]{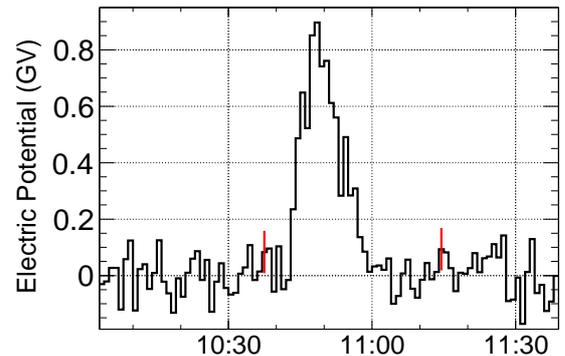}
%\vskip -0.1in
\caption{\label{fig05} Estimated electric potential shows a
                       maximum of (0.90\,$\pm$\,0.08)\,GV at
                       10:48\,UT on 1 December 2014. Vertical
                       bars represent $\pm$\,1$\sigma$ error.}
\end{center}
\vskip -0.2in
\end{figure}
Thunderclouds are known to have a complex multipolar structure
\cite{Mason72}, but here it is assumed to be dipolar since the
implications of such a structure can be easily simulated and a
quantitative comparison of simulation output with experimental
data could be used to obtain the average properties of the
thundercloud by treating it as a parallel plate capacitor that
can only provide an approximate estimate of its properties. To
simulate muon response to thundercloud potential {\bf V}, a
uniform vertical electric field {\bf E$_i$} for the following
three cloud thicknesses {\bf D$_i$} were investigated, where
{\bf V\,=\,E$_i$\,D$_i$}. (1) {\bf D$_1$}\,=\,2\,km for field
between 8 and 10\,km amsl, (2) {\bf D$_2$}\,=\,7.8\,km for
field between the ground and 10\,km amsl, (3)
{\bf D$_3$}\,=\,10\,km for field between 10 and 20\,km amsl.
The dependence of $\Delta$I$_{\mu}$ on {\bf V} was obtained
from Monte Carlo simulations, described in the next paragraph
and was found to be same for cases (1) and (2). For case (3)
$\Delta$I$_{\mu}$ was 15\% smaller than cases (1) and (2).
Thus, the case (3) apart from being unrealistic, also
required potentials higher than other two cases. Thus, a
uniform electric field applied between 8 and 10\,km was used
to provide a conservative estimate of the thundercloud
potential {\bf V}.

\begin{figure}[h]
%\vskip -0.05in
\begin{center}
\includegraphics*[width=0.4\textwidth,angle=0,clip]{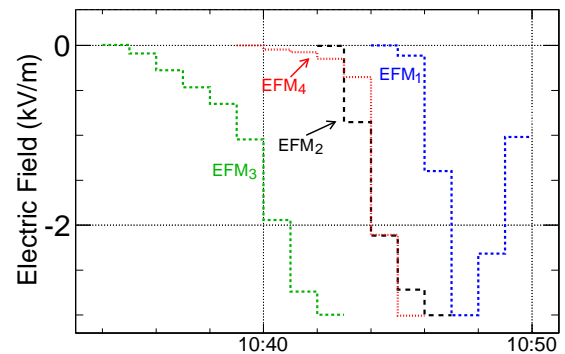}
\vskip -0.1in
\caption{\label{fig06} EFM$_3$ profile appears first, followed
                       by EFM$_2$ and EFM$_4$ after a 4\,min
                       delay. EFM$_1$ comes last, 6\,min after
                       EFM$_3$. Based on these EFM delays and
                       locations from Fig.\,\ref{fig01}c, a
                       thundercloud velocity of 1\,km\,min$^{-1}$
                       from east to west shown schematically in
                       Fig.\,\ref{fig01}d is inferred.}
\end{center}
\vskip -0.4in
\end{figure}
\begin{figure*}[t]
\begin{center}
\includegraphics*[width=0.9\textwidth,angle=0,clip]{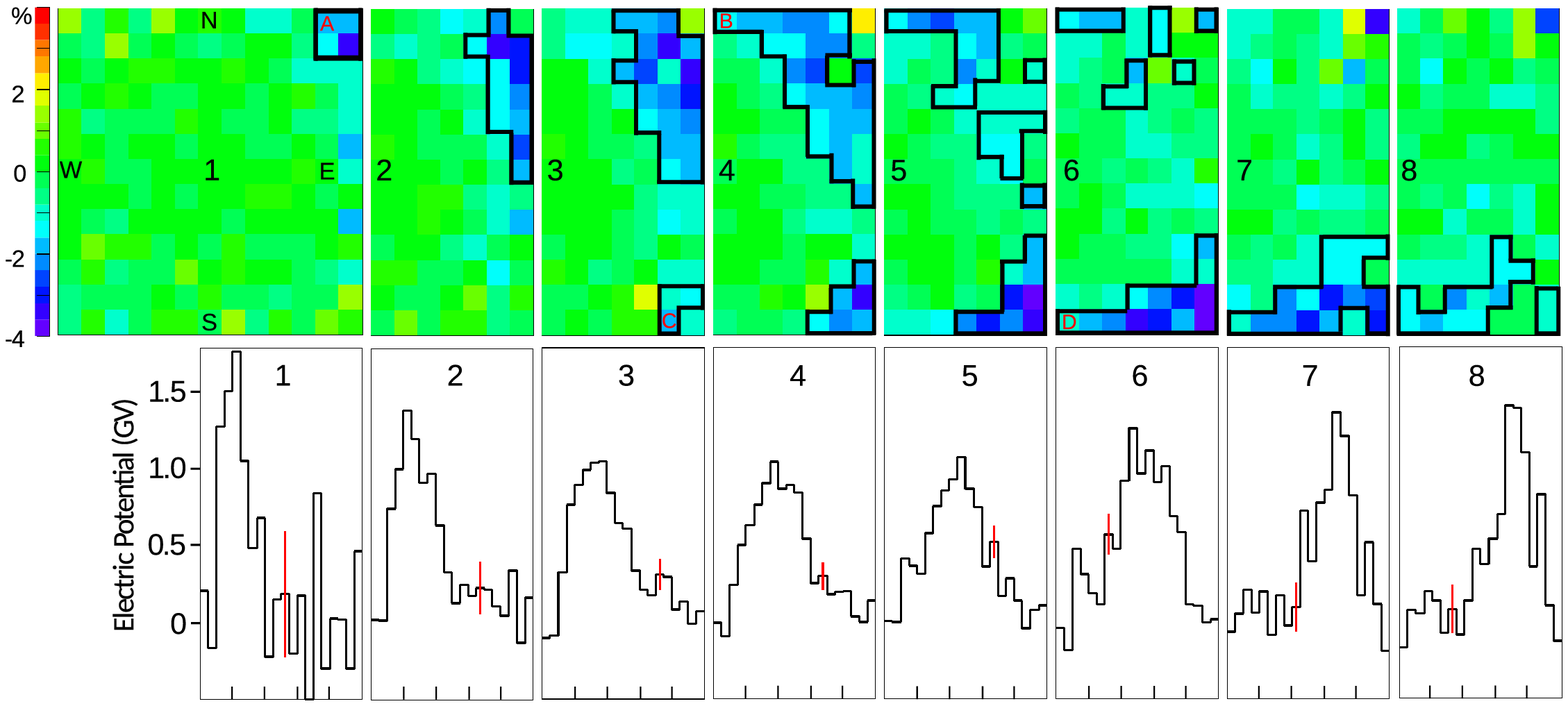}
\vskip -0.1in
\caption{\label{fig07} Top 8 panels show affected directions for
                       successive 2\,min exposures starting 1
                       December 2014 10:42\,UT. Bottom 8 panels
                       show estimated potentials needed to
                       reproduce $\Delta$I$_{\mu}$ shown in
                       the corresponding panel above for a
                       20\,min duration (10:41--11:00\,UT).
                       Maximum potentials of 1.8, 1.4, 1, 1,
                       1.1, 1.2, 1.3, 1.4\,GV (mean\,=\,1.3\,GV)
                       observed for panels, 1 through 8. Angular
                       velocity of 6.2$^{\circ}$min$^{-1}$,
                       inferred for directions (i) {\bf A} to
                       {\bf B}, (ii) {\bf C} to {\bf D} in north
                       and south FOVs, respectively are shown in
                       Fig.\,\ref{fig01}d. Vertical bar in
                       each bottom panel corresponds to
                       $\pm$\,1$\sigma$ error.}
\end{center}
\vskip -0.35in
\end{figure*}
The conversion of observed $\Delta$I$_{\mu}$ into equivalent
potential {\bf V} is derived from Monte Carlo simulations
using the CORSIKA code \cite{CORSIKA}, that in turn relies
on the choice of hadronic interaction generators. Here, FLUKA
\cite{FLUKA} and SIBYLL \cite{SIBYLL} were used for the low-
($<$80\,GeV) and high-energy ($>$80\,GeV) interactions,
respectively. When two other popular high-energy generators,
namely, QGSJet \cite{QGS} or EPOS \cite{EPOS} were used, an
identical dependence of $\Delta$I$_{\mu}$ on {\bf V} was
obtained. This is because the affected muons are produced by
low-energy ($<$80\,GeV) CRs where the high-energy generators
are not used. But, when the other two low-energy generators,
GHEISHA \cite{GHEISHA} or URQMD \cite{URQMD} were used
significant differences were observed. Compared to FLUKA,
{\bf V} inferred for GHEISHA was on an average 15\% higher,
and for URQMD 6\% higher. FLUKA was chosen as it provided the
lowest and therefore, the most conservative estimate of the
thundercloud potential. Next, the Monte Carlo simulation of
muons detected by the G3MT in each of the 169 directions were
carried out, first with {\bf V}\,=\,0, and then by applying a
{\bf V} in the range --3\,GV to 3\,GV in 0.1\,GV steps over a
height from 8 to 10\,km amsl as explained above. For each
direction, the number of muons above the corresponding
threshold energy were calculated. A high-statistics muon
database of 10$^7$ for {\bf V}\,=\,0, and 10$^6$ muons for
each non-zero {\bf V} was created. This allowed the simulated
$\Delta$I$_{\mu}$ to be measured to 0.1\% accuracy, much
smaller than the error of 0.4--2.7\% in real data.

The solar-wind introduces a diurnal variation in I$_{\mu}$
that was removed by modeling with a higher-order polynomial
after excluding thunderstorm affected 18\,min data. The
change in I$_{\mu}$ during 18\,min is shown in
Fig.\,\ref{fig02}. A cluster of 45 contiguous directions
enclosed by dark-boundary displays significant decrease
in I$_{\mu}$ as shown in Fig.\,\ref{fig03}. During
10:42--10:59\,UT, a decrease of 2\% is visible 
to the right of the dark boundary in
Fig.\,\ref{fig02} with a 20\,$\sigma$ significance.

Simulated dependence of I$_{\mu}$ for 45 directions on
applied potential {\bf V} is shown in Fig.\,\ref{fig04}.
A positive {\bf V} at thundercloud top relative to bottom
would lead to energy-loss {\bf eV} for $\mu^{+}$ and gain
{\bf eV} for $\mu^{-}$. Since ratio
$\mu^{+}$/$\mu^{-}$\,$>$\,1.0, the loss of detected
$\mu^{+}$ exceeds the gain of $\mu^{-}$. Thus, the sum of
muons of both polarities decreases for positive {\bf V}
and beyond 1\,GV the slope gradually increases due to
rapid increase in decay probability of $\mu^{+}$ as seen
in Fig.\,\ref{fig04}. This dependence is used to convert
the measured $\Delta$I$_{\mu}$ into equivalent {\bf V}
that peaks at (0.90\,$\pm$\,0.08)\,GV as shown in
Fig.\,\ref{fig05}.

The EFM records of electric field (sample rate\,=\,20\,s$^{-1}$)
show a smooth profile with an r.m.s.\,=\,0.01\,kV\,m$^{-1}$ in
all four cases, same as the EFM resolution. This suggests the
absence of major lightening. Hereafter, mean electric field
(min$^{-1}$) is used for comparison with muon data (min$^{-1}$).
Since all EFM profiles were similar and their amplitudes varied
22\% around a mean\,=\,3.3\,kV\,m$^{-1}$, they were normalized
to 3\,kV\,m$^{-1}$ as shown in Fig.\,\ref{fig06}.
% a negative polarity implies an electric field pointing
% upward from the Earth. Despite their large separation the
% similarity of the EFM profiles indicates that the size of
% thundercloud was much larger than the maximum EFM
% separation of 6\,km.
EFM$_3$ after a delay of 4\,min was followed by EFM$_2$ and
EFM$_4$, both of which overlapped. EFM$_1$, closest to G3MT
was delayed by 6\,min relative to EFM$_3$, indicating a
thundercloud velocity of $\sim$1\,km\,min$^{-1}$, moving
from EFM$_3$ toward EFM$_1$ as shown schematically in
Fig.\,\ref{fig01}d.

Thundercloud movement in FOV may be studied by the displacement
of its muon image ($\Delta$I$_{\mu}$) in short 2\,min exposures.
Because short exposures reduce muon statistics thus, regions
that showed (I$_{\mu}$) decrease, (a) in contiguous directions
or, (b) isolated directions over $\geq$2 successive exposures
were selected. In Fig.\,\ref{fig07}, $\Delta$I$_{\mu}$ for first
exposure starting 10:42\,UT is shown for full FOV in first
top-panel labeled {\bf 1}. A decrease in 4 directions enclosed
by dark-boundary is visible, the potential needed is shown in
bottom-panel {\bf 1} of Fig.\,\ref{fig07} that shows a maximum
{\bf V}\,=\,1.8\,GV during 10:41--11:00\,UT. From the second
panel onward, only 91 affected directions in the east are
displayed. In top-panel {\bf 2}, 12 affected directions require
a maximum {\bf V}\,=\,1.4\,GV. This decreases to 1\,GV for panels
{\bf 3} (23) and {\bf 4} (32). Then it increases to 1.1\,GV and
1.2\,GV for panels {\bf 5} (28) and {\bf 6} (23), respectively.
Finally, reaches 1.4\,GV for panels {\bf 7} (16) and {\bf 8}
(13). Integer values in the parenthesis next to each panel-number
indicate the number of affected directions, highlighted by
dark-boundary in the corresponding top-panels.

Successive panels in Fig.\,\ref{fig07} show the west boundary
of the muon image moving from east-to-west in north-FOV. For
example, it moved from direction {\bf A} in top-panel {\bf 1}
to {\bf B} in top-panel {\bf 4} in 6\,min implying an angular
velocity of 6.2$^{\circ}$min$^{-1}$ as depicted in
Fig.\,\ref{fig01}d. A movement of 6.2$^{\circ}$min$^{-1}$ of the
% 37.3 deg in 6 minutes
muon image is seen in south-FOV from {\bf C} to {\bf D} in
top-panels {\bf 3} and {\bf 6}, respectively. A similar
movement  is also reflected in the progressive shift of peak
voltage in the eight bottom-panels of Fig.\,\ref{fig07}.
If this angular velocity (6.2$^{\circ}$min$^{-1}$) is combined
with linear velocity (1\,km\,min$^{-1}$) from EFMs, then a
height of 11.4\,km amsl is obtained, comparable to typical
thundercloud height (12\,km) \cite{Mason72}.
The 1\,km\,min$^{-1}$ velocity and 11.4\,km height is
consistent with the velocity and height of subtropical jet
stream in south India \cite{BritEncl}.

In north-south direction the muon image covers the full FOV
that corresponds to an angular size of 74.6$^{\circ}$ as seen
in Fig.\,\ref{fig02}. This implies a radius of $\geq$11\,km,
very similar to average thundercloud radius ($\sim$12\,km)
\cite{NOAA} and yields total area of this thundercloud of
$\geq$380\,km$^2$. A thundercloud with infinitesimally thin
charged regions, separated by 2\,km acts as a parallel-plate
capacitor of capacitance $\geq$1.7$\mu$F. But in reality
thickness of charged regions is comparable to their
separation that reduces capacitance by $\sim$50\% to
$\geq$0.85$\mu$F. {\bf V}\,=\,1.3\,GV would require total
charge Q\,=\,$\geq$1100\,Coulomb and energy of
$\geq$720\,GJ stored in this thundercloud. A 1.3\,GV
potential across the thundercloud with its two charged
regions of thickness 2\,km each and a distance of 2\,km
between them implies an average field of 2.2\,kV\,cm$^{-1}$
which is lower than the breakdown field at high altitudes
\cite{Mason72}. The mean time to reach the maximum
potential shown in eight bottom panels in Fig.\,\ref{fig07}
is 6\,min. Thus, the thundercloud would have delivered a
power of $\geq$2\,GW, comparable to single biggest nuclear
reactors \cite{Pioro15}, hydroelectric and thermal power
generators \cite{Power}. Separation of 2\,km used is
reasonable since it extends the thundercloud top into
tropopause that defines the limit of cumulonimbus clouds
producing major thunderstorms in the atmosphere
\cite{Mason72}. Since the capacitance, total charge,
energy stored and power delivered by a thundercloud vary
inversely with the separation of its charged layers, thus
these parameters can be easily calculated for any other
separation.

The potential can be measured by integrating electric
field over thundercloud height. However, in general the
field measured by instruments aboard aircraft and
balloons span a region much smaller than the
thundercloud height and therefore, can not provide a
reliable estimate of the potential. On the other hand,
the parameter $\Delta$I$_{\mu}$ depends on the
thundercloud potential and is virtually independent of
its electric field and/or height. This makes muon
telescopes with GeV threshold such as the G3MT ideal
for measuring GV potentials in thunderclouds. However,
such high-potentials can not be indefinitely sustained
and a breakdown of air would result in acceleration of
electrons to GeV energies. It is conceivable that
bremsstrahlung emission from GeV electrons could
produce photons ranging from a few to beyond 100\,MeV
in a short flash of terrestrial $\gamma$-rays.

%\noindent 
{\it Conclusions.---}
The GRAPES-3 muon telescope is well-suited to measure the
electric potential developed in thunderclouds as shown for
the 1 December 2014 event where a peak electric potential
of 1.3\,GV was measured. This value is an order of
magnitude larger than the previously reported maximum of
0.13\,GV. This possibly is the first direct evidence for
the generation of GV potentials in thunderclouds,
consistent with the prediction of C.T.R. Wilson, 90 years
ago \cite{Wilson29}. The existence of GV potentials could
explain the production of highest energy $\gamma$-rays in
terrestrial $\gamma$-ray flashes discovered 25 years back
\cite{Fishman94}. It is shown that a $\geq$2\,GW of power,
comparable to single biggest nuclear reactors
\cite{Pioro15}, hydroelectric and thermal power generators
\cite{Power} was delivered by this thunderstorm that was
estimated to be moving at speed of 60\,km\,h$^{-1}$ near
the top of the troposphere. Despite a simplified
structure of the thundercloud used here, the present
work provides reasonable insights into the physical
state of the thunderstorms.

\begin{acknowledgments}
\vskip -0.15in
D.B. Arjunan, V. Jeyakumar, S. Kingston, K. Manjunath, S.
Murugapandian, S. Pandurangan, B. Rajesh, K. Ramadass, V.
Santhoshkumar, M.S. Shareef, C. Shobana, R. Sureshkumar are
thanked for assistance in running the GRAPES-3 experiment.
The GRAPES-3 experiment was built with generous support of
TIFR and the department of atomic energy, government of
India. This work was partially supported by the grants
from ISEE, Nagoya University, the Chubu University, and
the Ministry of Education and Science, Japan. We thank
the three anonymous referees whose prompt, critical and
constructive comments led to a significant improvement in
the final manuscript and its early publication.
\end{acknowledgments}


\begin{thebibliography}{99}

\bibitem{Franklin50}
%https://founders.archives.gov/documents/Franklin/01-04-02-0006;
B. Franklin, Experiments and Observations on Electricity made at Philadelphia
in America (London, 1751); B. Franklin, Phil. Trans. {\bf 47}, 565 (1752).

\bibitem{Wilson29}
C.T.R. Wilson, Nucl. J. Franklin Inst. {\bf 208}, 1 (1929); Proc. Phys. Soc. Lond. {\bf 37}, 32D (1924); Proc. R. Soc. Lond. A {\bf 236}, 297 (1956).


\bibitem{Mason72}
B.J. Mason, Proc. R. Soc. Lond. A {\bf 327}, 433 (1972); ibid {\bf 415}, 303 (1988); J. Mason and N. Mason, Eur. J. Phys. {\bf 24}, S99 (2003); E.R. Williams, Sci. Am. {\bf 259}, 88 (1988); C.P.R. Saunders, Space Sci. Rev. {\bf 137}, 335 (2008).

\bibitem{Fishman94}
G.J. Fishman et. al, Science {\bf 264}, 1313 (1994).

%\bibitem{Smith05}
%D.M. Smith, L.I. Lopez, R.P. Lin and C.P. Barrington-Leigh, Science {\bf 307}, 1085 (2005).

%\bibitem{Briggs10}
%M.S. Briggs et. al, J. Geophys. Res. {\bf 115}, A07323 (2010).

\bibitem{Tavani11}
M. Tavani et. al, Phy. Rev. Lett. {\bf 106}, 018501 (2011).

\bibitem{Marshall01}
T.C. Marshall and M. Stolzenburg, J. Geophys. Res. {\bf 106}, 4757 (2001).

\bibitem{Dwyer03}
J.R. Dwyer et. al, Science {\bf 299}, 694 (2003).

\bibitem{Ringuette13}
R. Ringuette et. al, J. Geophys. Res. {\bf 118}, 7841 (2013).

\bibitem{Alexeenko87}
%V.V. Alexeenko, A.B.Chernyaev, A.E. Chudakov, N.S. Khaerdinov, S.K. Ozrokov and% V.G. Sborshikov, Proc. Int. Cosmic Ray Conf. {\bf 4}, 272 (1987).
V.V. Alexeenko et al., Proc. 20th International Cosmic Ray Conf. {\bf 4}, 272 (1987).

\bibitem{Dorman03}
L.I. Dorman et. al, J. Geophys. Res. {\bf 108}, 1181 (2003).
%DOI:10.1029/2002JA009533

\bibitem{Muraki04}
Y. Muraki et al. Phys. Rev. D {\bf 69}, 123010 (2004).

\bibitem{Chilingarian17}
A. Chilingarian et al. Sci. Rep. {\bf 7}, 1371 (2017).

\bibitem{Hari17}
B. Hariharan et al. Proc. Sci. PoS (ICRC2017) 481.

\bibitem{Gupta05}
S.K. Gupta et al. Nucl. Instrum. Methods A {\bf 540}, 311 (2005); Y. Hayashi et 
al. Nucl. Instrum. Methods A {\bf 545}, 643 (2005).%; H. Kojima et al. Phys. Rev. D {\bf 98}, 022004 (2018).

\bibitem{Mohanty16}
P.K. Mohanty et al. Phys. Rev. Lett. {\bf 117}, 171101 (2016); P.K. Mohanty et al. Phys. Rev. D {\bf 97}, 082001 (2018).

\bibitem{Boltek}
https://www.boltek.com/EFM-100C\_Manual\_121415.pdf

\bibitem{CORSIKA}
https://www.ikp.kit.edu/corsika

\bibitem{FLUKA}
http://www.fluka.org/references.html

\bibitem{SIBYLL}
E.J. Ahn, R. Engel, T.K. Gaisser, P. Lipari, and T. Stanev, Phys. Rev. D {\bf 80}, 094003 (2009).

\bibitem{QGS}
N.N. Kalmykov, S.S. Ostapchenko, and A.I. Pavlov, Nucl. Phys. B Proc. Suppl. {\bf 52B}, 17 (1997).

\bibitem{EPOS}
T. Pierog et al., arXiv:1306.0121 [hep-ph] (2013).

\bibitem{GHEISHA}
http://cds.cern.ch/record/162911/files/CM-P00055931.pdf

\bibitem{URQMD}
http://urqmd.org

\bibitem{BritEncl}
https://www.britannica.com/science/subtropical-jet-stream

\bibitem{NOAA}
http://www.nssl.noaa.gov/primer/tstorm/tst\_basics.html

\bibitem{Pioro15}
I. Pioro and R. Duffey, ASME J. of Nucl. Rad. Sci. {\bf 1}, 011001 (2015).

\bibitem{Power}
%https://water.usgs.gov/edu/hybiggest.html
https://www.power-technology.com/features/feature-the-10-biggest-hydroelectric-power-plants-in-the-world; https://www.power-technology.com/features/feature-giga-projects-the-worlds-biggest-thermal-power-plants.

\end{thebibliography}
\end{document}